# “The tail wags the dog”: A study of anomaly detection in commercial application performance

Richard Gow[1,2], Srikumar Venugopal[1,2]
[1]School of Computer Science and Engineering
The University of New South Wales
Sydney, Australia
richard.gow@iag.com.au, srikumarv@cse.unsw.edu.au

Pradeep Kumar Ray[2]
[2]Asia Pacific ubiquitous Healthcare Research Centre
(APuHC), The University of New South Wales
Sydney, Australia
p.ray@unsw.edu.au

*Abstract*— **The IT industry needs systems management models that leverage available application information to detect quality of service, scalability and health of service. Ideally this technique would be common for varying application types with different n-tier architectures under normal production conditions of varying load, user session traffic, transaction type, transaction mix, and hosting environment.**

**This paper shows that a whole of service measurement paradigm utilizing a black box M/M/1 queuing model and auto regression curve fitting of the associated CDF are an accurate model to characterize system performance signatures. This modeling method is also used to detect application slow down events. The technique was shown to work for a diverse range of workloads ranging from 76 Tx/ 5min to 19,025 Tx/ 5min. The method did not rely on customizations specific to the n-tier architecture of the systems being analyzed and so the performance anomaly detection technique was shown to be platform and configuration agnostic.**

## I. Introduction

Commercial system management is enhanced via simplification and standardization of people training, processes, and technology solutions. It is therefore preferable to have a common measurement mechanism that can be used to manage availability SLAs (Service Level Agreements), performance SLAs, scalability OLAs, detection of anomalies and automated resource provisioning in the plethora of possible hosting environments available commercially. Accordingly, it is desirable to have a performance anomaly detection technique that is not specific to any particular platform or “n”-tier arrangement. This platform and configuration agnostic technique needs to prove itself capable of detecting performance anomalies on varying application types under normal production conditions of varying load, user session traffic, transaction type, transaction mix, and hosting environment.

Managing the stability and performance of the end-user experience using high level reliable information models is critical to enterprises and has a direct bearing on productivity and effectiveness of business processes. Most large enterprises have private data center arrangements with varying degrees of infrastructure rationalization. This creates a shared resource environment with the potential for unintended consequences as consolidation and sharing resources drive down Total Cost of Ownership. Collateral damage of one application service by another is common but poorly characterized across application portfolios. The ability to understand application performance and pro-actively manage their state is becoming increasingly important as infrastructure services move towards commoditization models such as cloud computing.

Pro-active systems management involves detecting abnormal performance situations that will result in a degradation of service via changes to application signatures. An application performance signature is driven by a complex mix of common resources – CPU, memory, I/O channels, worker threads, network conditions, storage arrangements, shared platforms, configurations and transaction mix. Characterizing all of these is difficult and requires detailed baseline calibration that needs to be re-done when the application, resources, or transaction mix change.

In this paper, we explore the construction of black box models that take a “whole of system view” and abstract the underlying complexity of the system being analyzed. The contribution of this paper is to introduce a simplified method of characterizing application performance signatures that recognizes the transaction tail and is not customized to the application being analyzed. This signature is characterized by the regression parameters of the service time cumulative distribution function (CDF) adapted from the general form of the M/M/1 queuing model. We demonstrate how detection of performance anomalies is achievable via tracking changes in these parameters using a probabilistic distribution of performance deviations between old and new conditions. The method is platform agnostic, and does not require extensive calibration and model re-work with changing conditions.

The remainder of this paper is structured as follows. Section II discusses related work in this area. Section III introduces the proposed performance signature and anomaly detection model. Section IV overviews the commercial systems analyzed and sampling techniques used. Section IV also presents experimental validation of the model introduced in III. Section V summarizes and discusses key results. Section VI discusses the limitations of the detection method and section VII presents the conclusions and future research ideas.

## II. Related Work

Past publications ([1], [3], [5], [7]) have discussed and showed the use of queuing theory laws and derivations, service time CDFs and regression techniques to characterize application performance. These have involved deriving application performance signatures and comparing them over time to determine performance anomalies.

Urgaonkar, et al. [7] discussed the problem of modeling multitier Internet applications using a network of queues with each tier represented by a queue or queues depending on the presence of load balancing. Sessions were modeled using an infinite server queuing system that fed the multitier queuing model and formed a closed queuing system. Stewart, Kelly and Zhang [13] used an M/M/1 queuing model for performance prediction in changing transaction mix workloads. They demonstrated that accounting for transaction mix non-stationarity enabled more accurate prediction of application-level performance in steady state compared to models that predict performance based on workload that ignores transaction type. Lama and Zhou [14] achieved efficient server provisioning using performance modeling via an M/GI/1/PS queuing model. The end-to-end service time $90^{th}$ percentile was used to model the need for more or less horizontal scaling in order to guarantee end-user response times. The queuing model approach involved low-level characterization of application components.

The above models use lightweight passive measurements routinely collected in production environments. However, the modeling methods used were complex and required extensive monitoring across all technology tiers and representative baseline runs to determine queuing parameters. They used various averaging techniques to represent the detail and complexity in the model - average service times; average queue visit ratios; average think times for user sessions; and average concurrent session loads - in order to calculate the average response times of requests. As discussed in Downey & Feitelson [2], averages are a blunt instrument when representing workload with long tailed service time distributions. Grade of service percentiles were suggested as a more appropriate way of representing service time distributions with "long tails".

Application performance issues have direct and immediate impact on end-user experience and hence satisfaction. Cherkasova et al [3] discussed online automatic detection of performance anomalies and application changes via integration of two complementary techniques. The first technique involved creating a resource consumption model of an application that, using regression, correlated processed transactions and consumed CPU time. The aim was to maintain a model that reflected resource consumption during normal operations. Significant changes in CPU resource consumption were then used to identify performance anomalies or application changes.

The second technique in Cherkasova, et al. [3] used the same concept of creating an application performance signature discussed in Mi, et al. [1]. This involved creating a compact run-time model of application behavior via derivations of queuing theory models - Little's Law and the Utilization Law. In this technique, a representative application server transaction service time was generated for all individual transaction types. This was done by creating a CDF for increasing service times across a range of server utilizations. Changes in the $50^{th}$ percentile of the CDF were used to characterize performance changes due to anomalies.

The use of regression as a technique to profile application performance was also discussed and used in Zhang, et al. [5] to project resource provisioning based on profiling CPU usage. The model accurately determined maximum achievable throughput for transaction mixes. It did this using non-negative Least Squares Regression to produce an approximation of the CPU processing cost of all transactions across all tiers. This was then input to the Mean-Value Analysis (MVA) algorithm [6] to compute the mean response time, average system throughput, and average queue length.

Urgaonkar, et al. [12] used the same MVA algorithm to identify bottlenecking and dynamically provision, predict response times, manage application configuration and session policing. Instead of using regression, the model used extensive technology tier monitors and representative baseline runs to determine queuing parameters. Model results were validated using synthetic workloads on real systems. The model was, however, challenged by nonstationary workloads. This work also assumed that key model parameters determined in low system utilisation were applicable during workloads being modelled. These limitations impacted the applicability of the models to real systems.

The methods discussed so far tend to be designed for specific application architectures. They require detailed baseline calibration that needs to be re-done whenever changes are made to the application resources and or transaction mix. They frequently use averaging techniques that are limited in describing an environment that is driven by changing transaction and workload mixes, and significantly impacted by the transaction tail. An application signature characterization and modeling approach is required that is application & architecture agnostic, and accounts for the tail.

Dean and Barroso [15] discuss concepts associated with the notion that "the tail" of transaction performance impacts the application service. The key insights were that even rare performance hiccup events affect a significant fraction of all requests in large-scale distributed systems. This is especially true with high workload arrival rates because shared resources such as infrastructure components and software worker threads can be quickly consumed by a small fraction of the arriving workload.

In this paper, the ability to pre-emptively detect degrading performance is best driven by characterizing changes in the "service tail" of the CDF exponential curve. This is the most sensitive part of the service time distribution. Failure of the application service is exponential in its performance loss if this "tail area" performance degrades. During anomalous performance events, the exponential service time distribution means that the tail area can dramatically impact the scalability of the system. This means that system behavior can be thought of as a case where "the tail wags the dog".

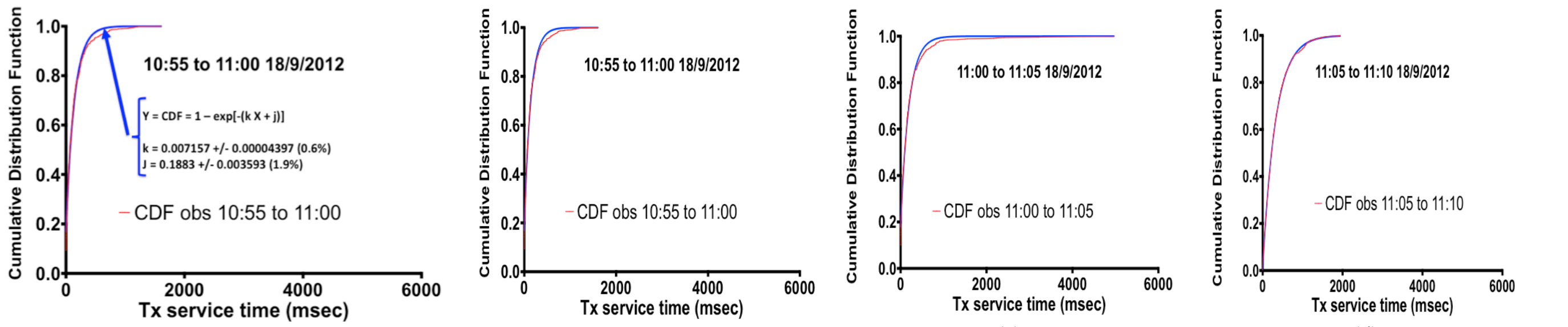

(a) (b) (c) (d)

Fig. 1: (a) CDF curve fitting example (b) – (d) Example performance anomaly detected on the online supply chain management system.

Detection and remediation are separate capabilities. Remediation to avoid service level breakdowns is the subject of current research ([8],[9],[10],[11]). This remediation is outside the scope of this paper.

## III. Proposed Detection Method

The technique presented in this paper seeks to use the concept of application performance signatures discussed previously. However, it was applied at a "whole of application service" level instead of building application service models from individual transaction type measurements, platform infrastructure profiles (e.g. CPU), transaction modeling and n-tier queuing models.

The service time CDF for online application service performance in an M/M/1 queuing theory model is an exponential curve as per equation (1) below. This is the most common type of queue and involves a single server model. It requires the transaction inter-arrival and service times to be exponentially distributed.

$$F(r) = 1 - e^{-r\mu(1-\rho)} \qquad (1)$$

$r$ = response time
$\mu$ = service rate of jobs per unit time
$\rho$ = traffic intensity = $\lambda / \mu$
$\lambda$ = arrival rate in jobs per unit time
$F(r)$ =CDF of $r$

The general form of this CDF equation is used in this paper as the basis of detecting performance anomalies. The chosen general form of the CDF equation is:

$$Y = 1 - e^{-(kX+j)} \qquad (2)$$

where

$Y$ = CDF of the probability of a transaction service time being less than $X$
$X$ = observed transaction service time in msec.
$k, j$ = regression parameters

The detection method uses univariate nonlinear parametric regression least squares analysis to calculate the parameters in equation (2). This curve fits the observed service time CDF of the application service performance for each sample period. This assumes a Poisson arrival pattern for transaction workload arriving at the application service and an exponential service time profile.

Analyzing the regression parameters for each sample of transactions and tracking them over time aims to achieve the detection of application performance changes. This approach can be applied at the whole of system level as it only requires end-to-end transaction measurement at the front of the system being analyzed.

Fig. 1(a) shows an example of the observed service time CDF and the curve fitted regression parameters required to curve fit the general form of equation (2) for sample data obtained from an online supply chain management system in a large enterprise in Australia. The sample data was gathered every 5 min from 10:30am to midday on 18/9/2012.

The key challenge in determining what constitutes a reasonable performance anomaly detection capability is the balanced mix between sensitivity to changing conditions that may indicate a degrading performance situation and accuracy of predictions used in this capability. The balancing act is between being *too sensitive and too inaccurate* versus being *too insensitive and more accurate*. The idea is to detect degraded performance states rather than achieve absolute accuracy.

The sensitivity of this detection model was achieved because the curve fitting via regression characterizes the "service tail" of the CDF exponential curve. This was the most sensitive part of the service time distribution. Failure of the application service is exponential in its performance loss if this "tail area" performance degrades. Fig. 1(b)-(d) shows changing CDF curve shape during a performance anomaly event for the online supply chain management system.

An example of the regression parameter trending associated with this method is shown in Fig. 2(a) and (b) for $k$ and $j$ for the 5 min sample data from 10:30 to midday on 18/09/2012. This included the performance anomaly that is highlighted in the red box and which corresponds to the event in Fig. 1(b)-(d).

In this paper, detection of performance anomalies was defined via an adaptation of the change profile concept used by Shen, et al. [4]. Here, performance changes were detected via a probabilistic distribution of performance deviations between old and new conditions.

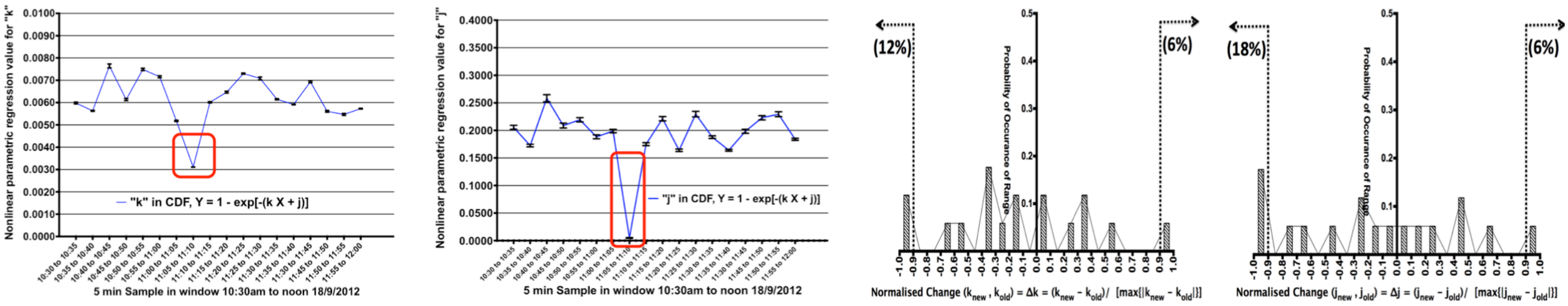

(a) (b) (c) (d)

Fig. 2: (a),(b) Example of regression parameter trending detecting a slow down event (c),(d) Visual representation of probability distribution of changes in regression parameters for the online supply chain management system.

$$Change(t_{old}, t_{new}) = (t_{new} - t_{old}) / max\{|(t_{new} - t_{old})|\} \quad (3)$$

where

$Change\ (t_{old}, t_{new})$ = degree of change in variable of interest ranging between -1 and +1

$t_{new}$ ,$t_{old}$ = values of change variable at new and old times respectively.

$max\{|(t_{new} - t_{old})|\}$ = normalize all –ve change values by dividing by the maximum –ve change. Similarly normalize all +ve changes.

This paper uses the application performance signature regression parameters that characterize each 5 min sample as the comparative metric between old and new conditions. The changes in $k$ and $j$ in Fig. 2 (a) and (b) vary between -1 and +1 based on the definition in Equation (3). The probability of outliers or large changes in $k$ can be determined from the probability distribution as shown in Fig. 2(c). As an example, if the 95% confidence level was chosen as the confidence level beyond which a change in $k$ was considered anomalous, the change in $k$ would need to be less probable than 0.05. This was approximately less than -0.92 or more than +0.9 in this particular sample. However, the changes in $k$ and $j$ were quantized into 0.1 increments and all events in this range were bundled into a "self similar" result set. The 5% significance level for self similar quantized delta sets requires that the extreme ends of the probability distribution were in groupings of less than or equal to 5%. Anything greater than this would result in the grouping being considered baseline normal behavior. Fig. 2 (c) shows that the most peripheral positive and negative groupings had a probability of 5.88% and 11.76% respectively. This meant that the significance level would need to be 6% or 12% respectively for these extreme scenarios to be considered anomalous. Increased sample size or a rolling probability distribution over an extended period may have addressed this issue by better defining the groupings.

However, the method could be further refined via a continuous distribution but simplification can be achieved via adjustment of the significance level to match the nature of the system being analyzed.

The problem with the probability distributions presented in Fig. 2 (c) and (d) was that the changes in $k$ and $j$ were widely spread across the range of possible values from -1 to +1. This indicated a system with variable performance or a sample too small to characterize behavior. A tightly bound grouping may have indicated an application that was more fine tuned and subject to little variation in performance.

Another important aspect of the detection model is to minimize processing required to produce a reasonable regression estimate. Consuming excessive resources and or taking too long to process detections is potentially uneconomic, and may inhibit the very pre-emptive ability being implemented. Auto regression engines are available commercially and have efficient data processing capabilities. Choosing an engine is outside the scope of this paper.

## IV. Analysis of Commercial Production Systems

A key aim of this research was to identify a performance anomaly detection technique that was platform and configuration agnostic. This technique needs to prove itself capable of detecting performance anomalies on varying application types under normal production conditions of varying load, user session traffic, transaction type, transaction mix, and hosting environment. Accordingly, this paper presents anomaly detection results from two different application types involving 2 and 3-tier configurations with different database arrangements, transaction types & mixes, and external service integrations. If a bottom up "n-tier" queuing model were created to establish normal performance behavior under varying workloads, each model would be unique. The overhead in running the various models would be significant. The example systems analyzed were commercially sensitive, and hence details are kept to a minimum in this paper.

### A. *Measurement & Sampling Techniques*

The systems exist in a private data center belonging to a large financial services enterprise primarily composed of IBM mainframe and mid-range platforms, shared among different applications across CPU & memory pools and SAN storage. Capacity was well provisioned but collateral impacts for short periods of time were common, especially during peaks.

Transaction information was collected in 5-minute samples. Transaction data was written to application service logs for each application server. Each sample was processed to produce summary application Grade of Service (GoS) performance data – transaction arrival rate, $50^{th}$ , $80^{th}$, $90^{th}$, $95^{th}$, $98^{th}$ and $100^{th}$ percentiles. Data was collected for Monday 4/2/2013 from 8am to 6pm for both applications. Monday was chosen because it was known to be the busiest day of the week.

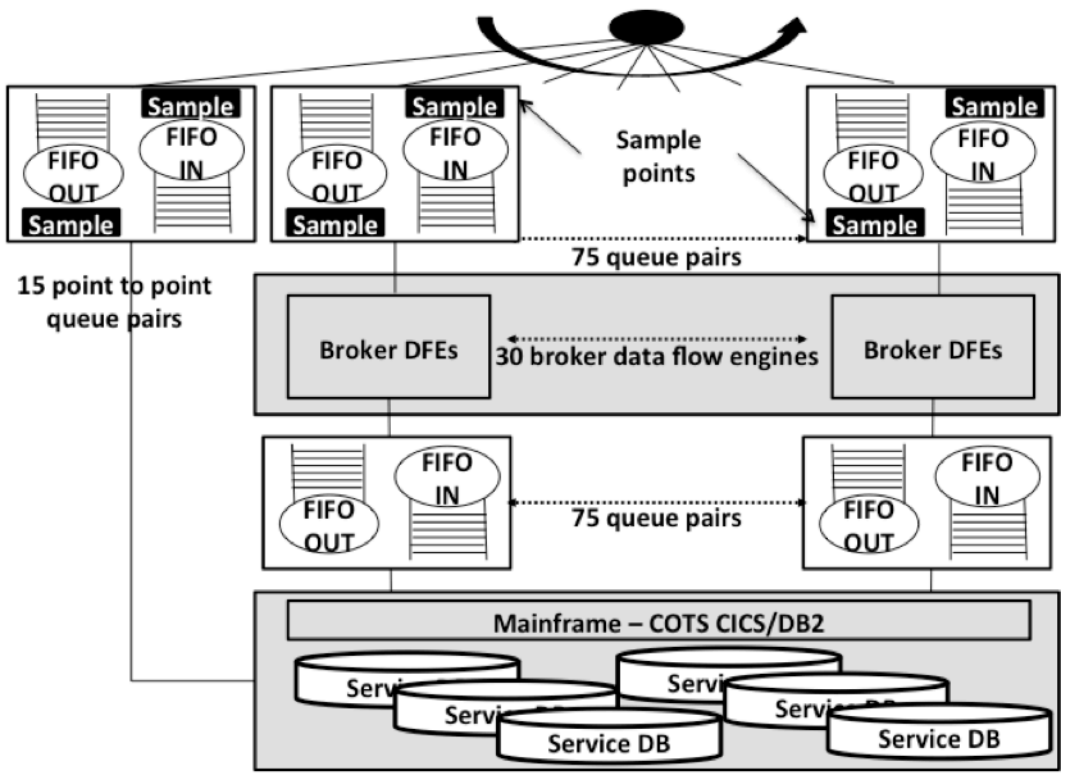

Fig. 3: High level architecture of mainframe system

TABLE 1: TRANSACTION PROFILE FOR MAINFRAME SYSTEM

| | | | | | |
|---|---|---|---|---|---|
| ***No. Txs in sample*** | | 1,664,010 | | | |
| ***No. Tx Types*** | | 206 | | | |
| ***Top 10 Transactions*** | | 48% of traffic | | | |
| ***No. Txs that were 1% or more of traffic*** | | 28 transactions | | | |
| **Traffic mix** | | | | | |
| ***Top 5 Txs (35.6% of all traffic)*** | Top Tx | $2^{nd}$ Tx | $3^{rd}$ Tx | $4^{th}$ Tx | $5^{th}$ Tx |
| | 14.9% | 7.7% | 5.1% | 4.8% | 3.1% |
| ***No. Tx Types in key workload percentiles*** | | $80^{th}$ | $90^{th}$ | $95^{th}$ | $100^{th}$ |
| | | 36 | 56 | 76 | 206 |

The detection method identified whether sample periods were anomalous. The method of validating whether the anomaly detection was correct or not involved checking the grade of service (GoS) performance data for the sample period and identifying if the performance percentiles showed abnormal slowdown. Verified events showed a slow down in some or all of the $50^{th}$ , $80^{th}$, $90^{th}$, $95^{th}$ , $98^{th}$ or $100^{th}$ percentiles.

Performance SLAs focused on the $95^{th}$ %ile. Detecting events in samples that showed movement in these six %iles was of interest to ensure SLAs were pro-actively protected. Investigation of root causes for these slowdowns was pursued and captured where identified.

### *B. Production Example 1 – Mainframe Product system*

The first system is an application service running on a mainframe system termed as the "Mainframe Product System". The high-level architecture of this application is illustrated in Fig 3. This application managed the product processes supporting all distribution and servicing channels in the Australian financial enterprise. The typical workload profile for this system is approx. 16-19,000 transactions per 5 minutes with 40 concurrent transactions on average and 100 maximum concurrent transactions during each weekday 10 am to 2pm peak. The transaction mix was nonstationary as it was driven by market conditions and product events.

Transactions submitted to this application were measured at the ingress and egress queues. Each transaction may have spawned many sub-transactions but the end-to-end transaction time was defined as the time from the initial ingress to the final egress. Table 1 shows the transaction mix for the 8am to 6pm sample on Monday 4/2/2013. This 10 hour sample contained approx. 1.7 million transactions in total with workload varying from 3,327 Tx/ 5 min up to a max of 19,025 Tx/ 5 min.

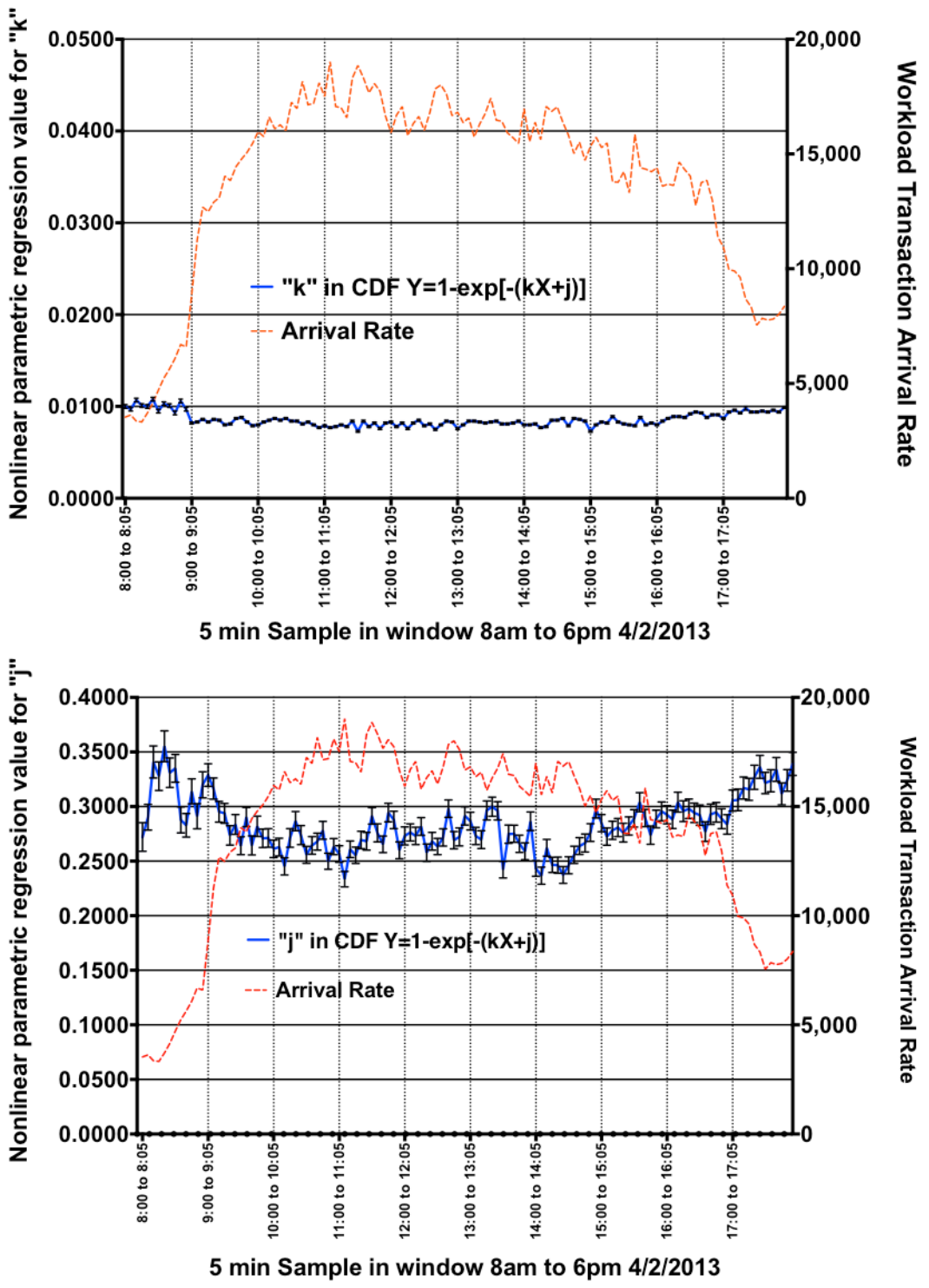

Fig. 4: Nonlinear parametric regression parameters *k* & *j* by time of day for Mainframe Product system.

The workload was concentrated in the top 10 transaction types (approx. 50%). The remaining 196 transaction types accounted for approx. 50% of the workload with only 28 transactions overall having 1% or more representation. The top transaction (at 14.9%) was a view product transaction from the core product source of truth M/F data store.

#### *1) Profiling changes in k and j*

Fig. 4 shows the time of day view of changes in *k* and *j*. There was an early morning decrease in *k* (system getting slower) as the workload increased. This stabilized once the transaction arrival rate reached 13,000-14,000 Tx/5 min and above around 9:30am. The workload peaked around 19,000 Tx/ 5min and moved back to 13-14,000 Tx/ 5min at the end of the day around 4:30pm. The value of k increased after 4:30pm indicating a performance increase. This suggested an inverse relationship between workload and *k*. Visually this relationship appeared to stabilize as the workload reached its normal peak bounds between 9:30am and 4:30pm and was possibly linear in this time. Its stable value was around 0.0083 (averaged over this period). The relationship between workload and parameter *k* may be able to be used to track application scalability over time and requires further research. There also appeared to be a similar time of day view of the relationship between parameter *j* and workload. It stabilized at 0.2768 (averaged over this period).

Fig. 5 (a) and (b) show the detection method results for the Mainframe Product System using a 5.0% significance level to select edge events. At the 5.0% significance level, all negative

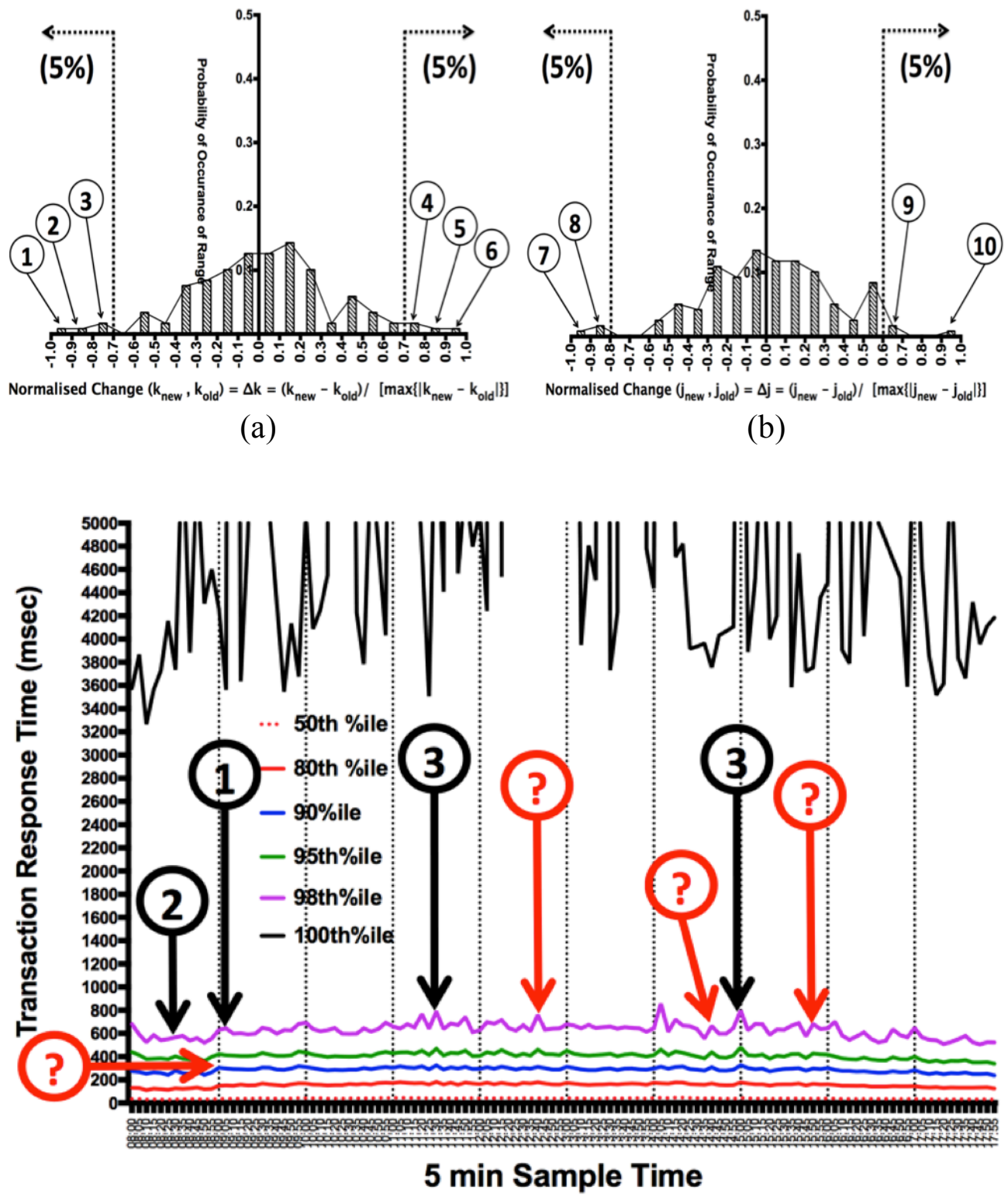

Fig. 5 : Probability distribution of changes in regression parameters (a) *k* and (b) *j* for mainframe application service. (c) Grade of service graph for the application's performance over time of day.

Events:

1) 09:00 Slow down all %iles except 100th %ile. Results in a -ve Δk offset by a +ve Δj
2) 08:30 slow down all %iles except 100th %ile. Results in a -ve Δk offset by a +ve Δj
3) 11:30 Slow down all %iles. Results in a -ve Δk offset by a +ve Δj
   15:00 Slow down all %iles. Results in a -ve Δk and a -ve Δj
4) 08:10 speed up in all %iles. Results in a +ve Δk and a +ve Δj
   15:45 Speed up all %iles. Results in a +ve Δk offset by a -ve Δj
5) 11:35 Speed up all %iles. Results in a +ve Δk offset by a -ve Δj
6) 08:50 Speed up all %iles. Results in a +ve Δk offset by a -ve Δj
7) 13:30 Slow down all %iles except 100th %ile. Results in a +ve Δk offset by a -ve Δj
8) 08:35 50th %ile stable. Speed up in 80,90,95th %iles. Slow down in 98th & 100th %iles. Results in a +ve Δk offset by a -ve Δj
   14:00 Slow down all %iles except 100th %ile. Results in a -ve Δk and a -ve Δj
9) 08:45 stable 5oth %ile, slow down in 80,90,95,98 & 100th %iles. Results in a -ve Δk offset by a +ve Δj
   13:35 Speed up all %iles except 100th %ile. Results in a +ve Δk and a +ve Δj
10) 08:10 speed up in all %iles. Results in a +ve Δk and a +ve Δj

Δ*k* events were missed. The changes in *k* and *j* were quantized into 0.1 increments and all events in this range bundled into a "self similar" result set. The 5% significance level for events meant that larger result sets which exceeded the 5% occurrence level were assumed to be baseline normal behavior e.g. -0.2 <= Δ*k* <= -0.1 in Fig. 5 (a).

The numbered events from 1 to 3 in the probability distribution in Fig. 5 (a) and (b) are marked on the Grade of Service (GoS) graph by time of day. The GoS graph showed several other small slow down events that might have reasonably qualified (see the question marks in Fig. 5(c)).

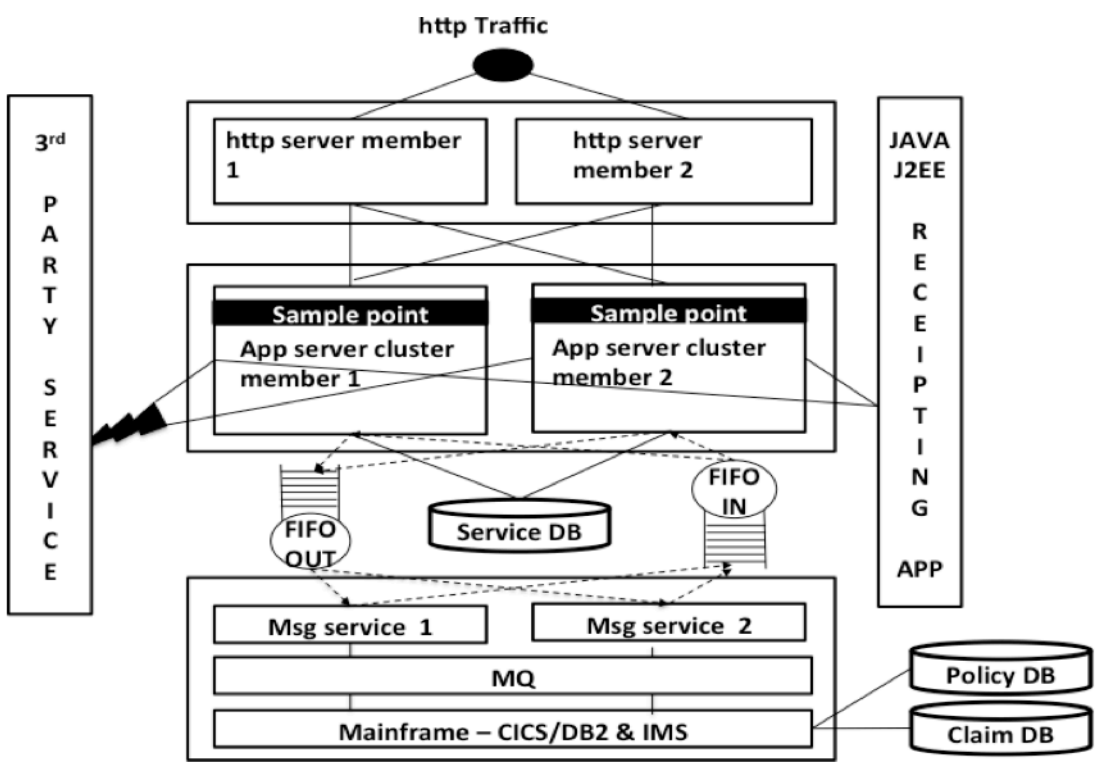

Fig. 6: High-level architecture of online payment processing system

TABLE 2: TRANSACTION PROFILE FOR PAYMENT PROCESSING SYSTEM

| | | | | | |
|---|---|---|---|---|---|
| ***No. Txs in sample*** | | 20,718 | | | |
| ***No. Tx Types*** | | 11 | | | |
| ***Top 10 Transactions*** | | 99.9% of traffic | | | |
| ***No. Txs that were 1% or more of traffic*** | | 8 transactions | | | |
| **Traffic mix** | | | | | |
| ***Top 5 Txs (89.1% of all traffic)*** | Top Tx | 2nd Tx | 3rd Tx | 4th Tx | 5th Tx |
| | 37.3% | 18.3% | 13.1% | 12.8% | 7.6% |
| ***No. Tx. Types in key workload percentiles*** | | 80th | 90th | 95th | 100th |
| | | 4 | 5 | 7 | 11 |

However, these were missed because of the 5.0% significance level and 0.1 quantization in the probability distribution. The choice of significance level needs to be small enough to avoid detection noise (false positives) yet sensitive enough to pickup missed events (false negatives). The aim of the detection method was to pickup true slow down events (true positives) and avoid false events (false positives and true negatives).

The detection method was specifically aimed at detecting slow down events rather than the speed up events that followed them. This required further understanding of the combinations of +ve and –ve changes in *k* and *j* that could be definitively associated with known slow down events in order to avoid false positives and false negatives from the detection method. This was very difficult in a low noise system such as the Mainframe Product System. Accordingly, the second system analyzed in Section IV C was chosen because its performance profile was known to be variable, and hence enabled detailed clarification of the combinations of *k* & *j* that could be definitively associated with slow down events. The objective was to avoid detection noise (false positives) even if this meant missing some smaller events (false negatives).

### *C. Production Example 2 – Payment processing system*

The second system analyzed managed credit card payments for all eBusiness, call center and face-to-face payment transactions for a large Australian financial services provider. Fig. 6 shows the high level architecture of this JAVA/ J2EE web application. It is a front-end web application clustered on 2 machines (members) for resilience. The application servicing entity was a JAVA/ J2EE application server hosted in the IBM WebSphere platform. HTTP transactions were associated with unique end-user session ids that were "sticky" because traffic traversed all higher infrastructure layers to and from the same application server instance. The driver of work for this front-end system was submissions from the end-user sessions that may have in turn spawned multiple sub-transactions to and from lower layers before returning the final result back to the end-user session. The total transaction response time was the time from receiving the end-user http request for service until the time the reply was sent back.

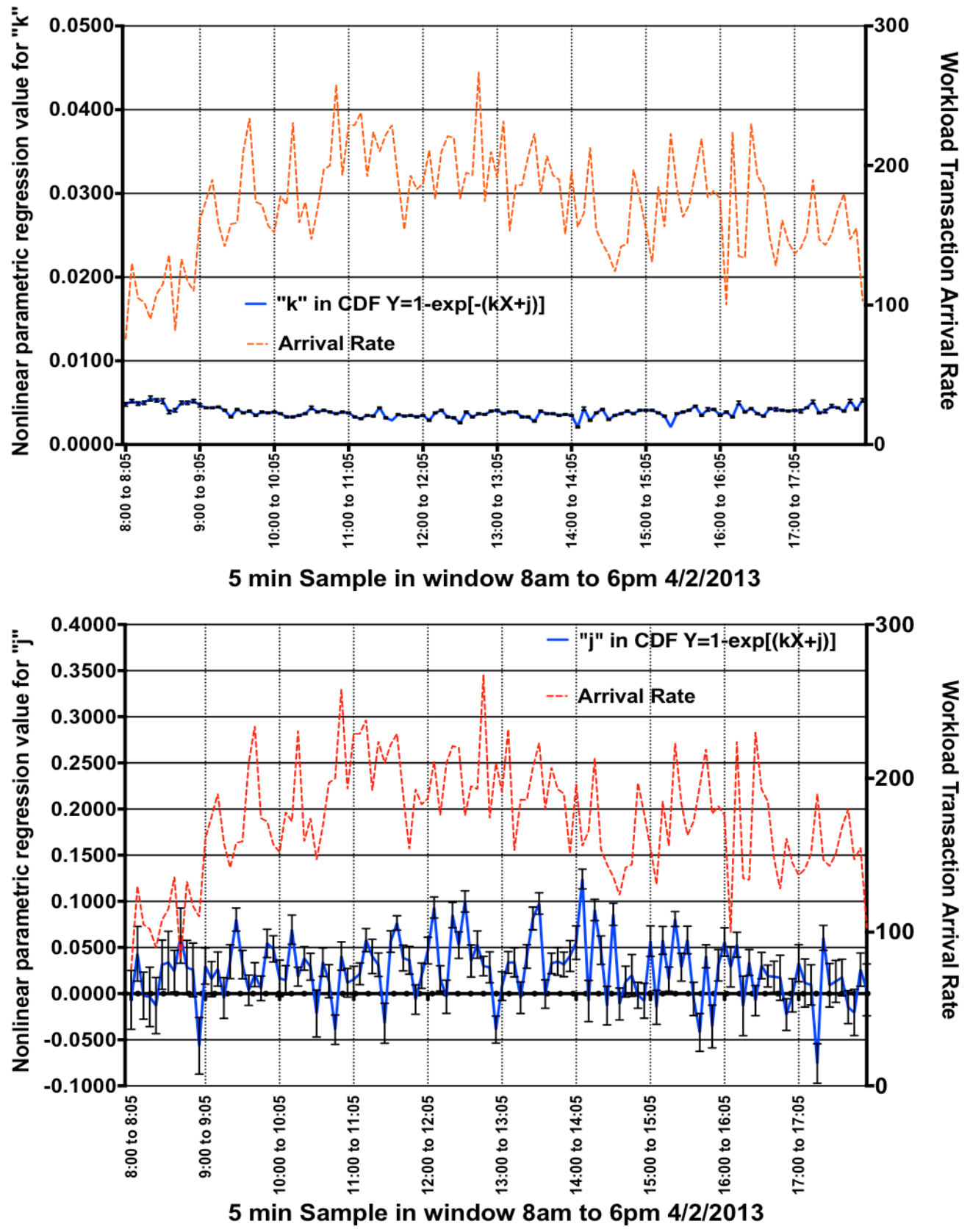

Fig. 7: Nonlinear parametric regression parameters *k* & *j* by time of day for payment processing system member 1

Sampling was performed on this system from 8am to 6pm on Monday 4/2/2013 every 5 minutes. During the sampling, the workload was found to have a peak arrival rate of 518 requests per 5 minutes with 112 peak active sessions per sample. On average, there were a max of 10 requests in the system concurrent in any sample going up to a maximum of 20. This workload was split evenly across the two cluster members.

The transaction mix was nonstationary as it was driven by retail behaviors that could be variable. Importantly, this application dealt with external payment gateway services shared amongst many large financial institutions and showed contention behaviors based on variable retail conditions.

*1) Payment processing system workload*

The application had two cluster members. They were both found to have the same results so only cluster member 1 is presented in this paper. Table 2 shows the transaction mix for this machine. This 10 hour sample contained approx. 21K transactions with workload varying from 76 Tx/5min up to a max of 267 Tx/5min. The workload was concentrated in the top 5 transaction types (approx. 90%). The remaining 6 transaction types accounted for approx. 10% of the workload with only 8 transactions overall having 1% or more representation. The top transaction (at 37.3%) was an external payments gateway access transaction. This heavy reliance on the external payments gateway service largely defined the transacting profile of this application service. Any Internet link or gateway transaction delays were a determining influence on performance.

*2) Profiling changes in k and j*

Fig. 7 shows the time of day view of changes in *k* and *j*. There was an early morning slight decrease in *k* (system getting slower) as the workload increased. This stabilized once the transaction arrival rate reached 200-300 Tx/5 min and above around 9:30am. At the end of the day, as the workload dropped below this point (around 5:30pm) the value of *k* slightly increased indicating a performance increase. Its stable value was around 0.0037 averaged over this period. The time of day view of *j* was quite variable about a linear trend. It averaged 0.0271 over this time. The amplitude of variation about this average and the error of the auto regression values were significantly larger than for the mainframe system discussed in IV B. It was also much more variable than the *k* parameter. This indicated instability in system performance compared to the low noise mainframe product system. Importantly, this instability was part of a broader problem subsequently diagnosed. Multiple mid-range systems all displayed instability to varying degrees when compared to the mainframe system. This was a key monitoring diagnostic used in preventing a severe organization wide mid-range outage. Additionally, it was found that the external gateway transactions for this application were being intermittently delayed via an under allocation of guaranteed bandwidth on the Internet link. Subsequent work doubled this allocation in order to reduce variable service times on these transactions.

The unique aspect of the payment processing system was its dependence on an external gateway service that issued tokens so that payment details could be masked in core systems. Setting up the external token issuing service and waiting for these tokens was synchronous in nature for individual user sessions. The service was performant but shared by many financial institutions such that small delay events for a variable number of sessions was a common feature of this application service transaction profile. These delays manifested as slow downs within the whole of service transaction mix followed by speed ups as the service freed up and completed issuing tokens.

The analysis approach chosen involved looking at the auto regression parameters for all 5min samples and analyzing approx. 20% of edge cases for +ve and -ve changes in *k* or *j* as measured by equation (3) to determine what type of performance change was detected. Fig. 8 shows the major grade of service change scenarios observed in this application service, and the resulting combinations of changes in *k* & *j* that resulted from curve fitting the CDF general form in equation (2).

The results from this application captured a representative sample of the combinations of grade of service slow downs and speed ups. This involved analyzing events in the sample

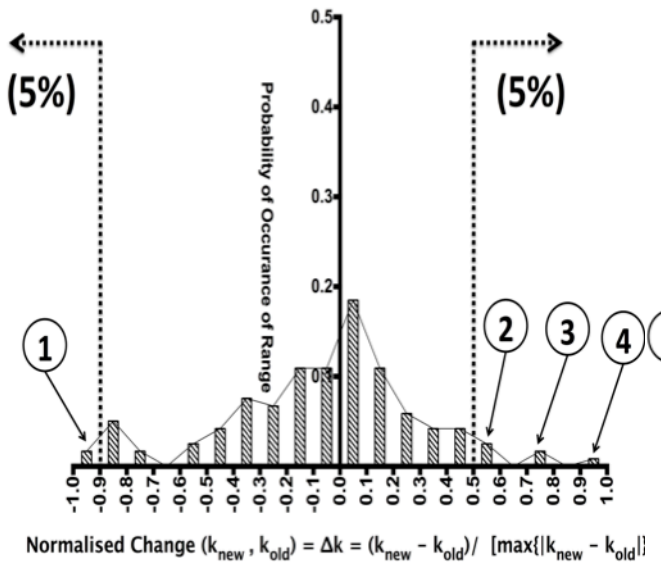

(a)

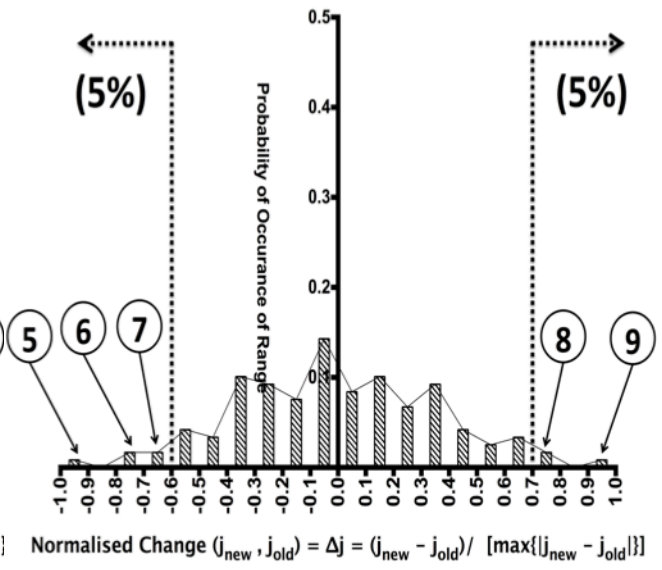

(b)

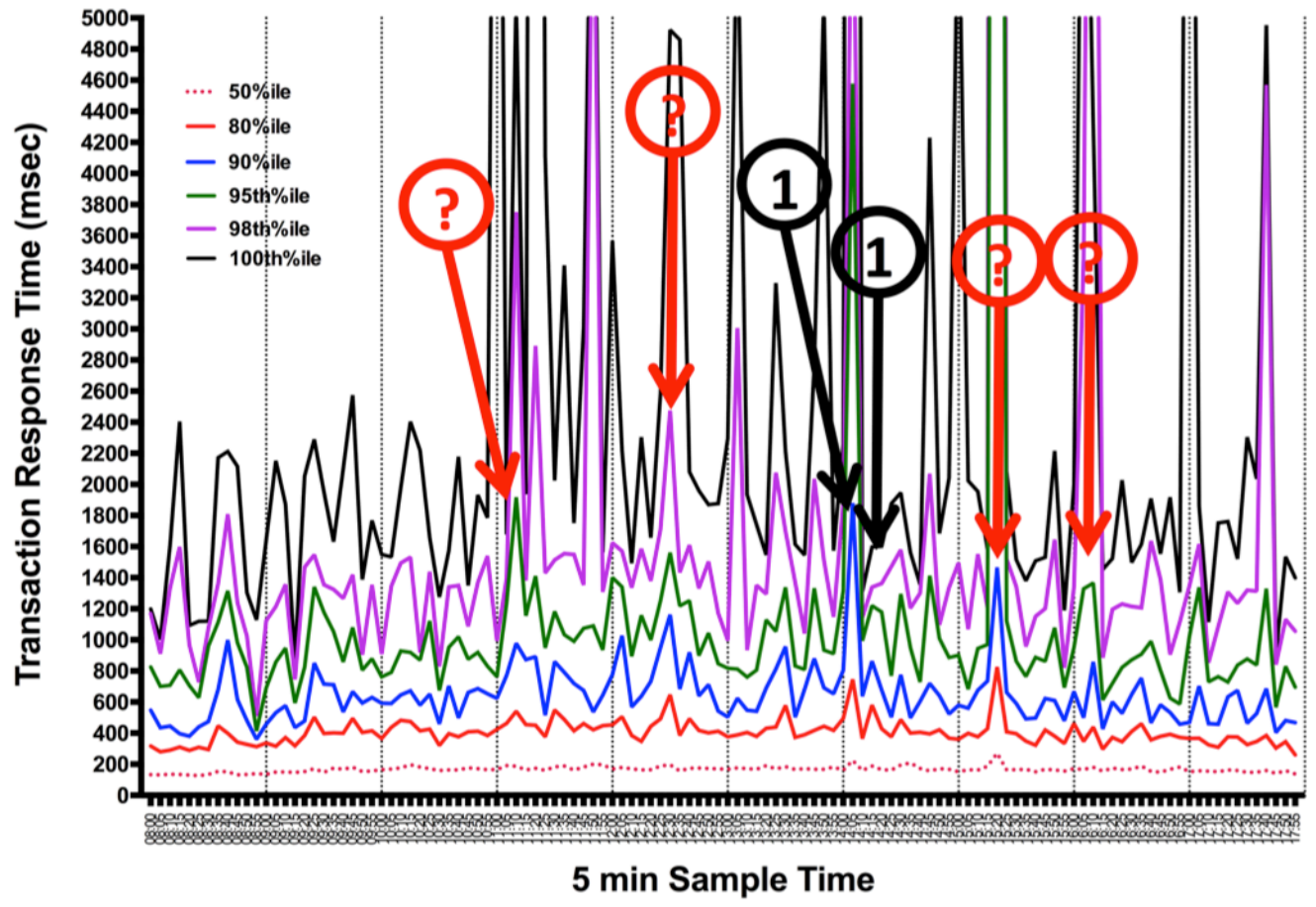

(c)

Fig. 8 : Probability distribution of changes in regression parameters (a) *k* and (b) *j* for payment processing service. (c) Grade of service graph for the application's performance over time of day.

Events:

1) 14:05 Slow down event all %iles causing a -ve Δ*k* offset by a +ve Δ*j*, 14:15 Slow down event all %iles causing a -ve Δ*k* offset by a +ve Δ*j*
2) 12:35 Speed up of all %iles. This causes a +ve Δ*k* offset by a -ve Δ*j*, 13:35 Speed up in all %iles results in a +ve Δ*k* offset by a -ve Δ*j* , 17:45 Speed up in all %iles results in a +ve Δ*k* and a -ve Δ*j* offset
3) 15:25 Speed up across all %iles following 15:20 slow down event causing a +ve Δ*k* offset by a -ve Δ*j*, 16:15 Speed up across all %iles causing a +ve Δ*k* offset by a -ve Δ*j*
4) 14:10 Significant speed up across all %iles following 14:05 slow down event causing a +ve Δ*k* offset by a -ve Δ*j*
5) 14:10 Significant speed up across all %iles following 14:05 slow down event causing a +ve Δ*k* offset by a -ve Δ*j*
6) 13:35 Speed up in all %iles results in a +ve Δ*k* offset by a -ve Δ*j*, 14:35 Speed up in all %iles results in a +ve Δ*k* offset by a -ve Δ*j*
7) 08:55 Speed up event all %iles results in a +ve Δ*k* and a -ve Δ*j* offset, 17:15 5oth %ile stable, speed up in 80 & 90th %ile, slow down in 95,98 & 100th %iles results in a +ve Δ*k* offset by a -ve Δ*j*
8) 14:15 Slow down event all %iles causing a -ve Δk offset by a +ve Δj, 14:30 Slow down event all %iles causing a -ve Δk offset by a +ve Δj
9) 17:20 Slow down 50,80,90,98 & 100th %iles, speed up in 95th %iles results in a -ve Δk offset by a +ve Δj

period. There were 4 combinations of +ve and –ve changes in the *k* and *j* auto regression parameters observed:

1. Speed up in main body and transaction tail
2. Hybrid scenario in which the transaction main body speeds up and the tail slows down
3. Slow down in main body and transaction tail
4. Hybrid scenario in which the transaction main body slows down and the tail speeds up

Significant slow downs resulted in a negative change in *k* where as speed-ups were characterized by a positive change in *k*. The changes in *j* could be +ve or –ve depending on the significance of the slow down or speed up event in *k* and hence changes in *j* were not a good predictor of event type.

Detailed results in the scenarios examined showed that a stable *k* with partial transaction slow downs involving the tail ($95^{th}$, $98^{th}$, and $100^{th}$ percentiles) were characterized by –ve changes in *j*. A stable *k* with speed-ups in the tail were characterized by +ve changes in *j*. Any changes in *k* could over ride these trends in *j*.

The results showed that *k* can be characterized as the "transaction main body" coarse grain adjustor in the general form of the CDF equation. The results showed *j* acted as the "transaction tail" fine grained adjustor. *k* was a determinant of event type. *j* determined event type only when *k* did not vary significantly. Positive changes were associated with speed up events and negative changes were associated with slow down events. There were 4 event scenario patterns:

1. Positive change in *k* = speed up event
2. Negative change in *k* = slow down event
3. Stable *k* plus positive change in *j* = speed up event in the transaction tail
4. Stable *k* plus negative change in *j* = slow down event in the transaction tail

The purpose of the detection method was to identify slow down events. This was most reliably determined via negative changes in *k* that were identified as anomalous by exceeding the change significance level, in this case 5%. The key issue with a highly variable application was that only considering the 5% significance level meant the detection method was insensitive to potential slow down events. This could be seen when mapping the slow down events from the probability distribution to the grade of service graph in Fig. 8(c). This meant two definite events were identified (annotation 1 in Fig. 8(a) and Fig. 8(c)) and four possible events missed for this sample of the payment processing system. The question mark annotations in the grade of service graph in Fig. 8(c) showed events that were missed due to the 5% significance choice.

The proposed detection method approach to identify slow down events was modified to the following:

1. Focus on events that involve negative *k* changes.
2. Quantize probability changes into 0.1 segments to allow grouping of events, simplify classification, and reduce false positive events
3. Set the significance level small enough to avoid false positives (5%)
4. Consider small changes to the significance level if the sample size means likely events are excluded by the 0.1 quantizing of probability segments.

Importantly, callouts 5, 6, and 7 in Fig. 8(b) couldn't be definitely characterized as they included both speed ups and slow downs. Negative changes in j could not be positively identified as slow downs.

## V. Discussion

As noted in the results discussed in Section IV C, the $k$ parameter mapped the early rise in the service time CDF and the $j$ parameter matched the tail of the curve. The $k$ parameter was a *coarse-grained transaction main body curve fitter* and the $j$ parameter was a *fine-grained transaction tail fitter.* Observing the variability in these regression parameters over time and between different applications was a good indicator of system issues. Extreme variability was found to be an effective indicator of several issues subsequently diagnosed:

a) Inadequate CPU allocation to applications and their I/O pools
b) Poor scheduling of batch style workloads such as dbase backups across all applications in shared resource pools
c) Problems with shared resource management in the operating system

The purpose of the detection method was to identify slow down events. This was most reliably determined via negative changes in $k$ that were identified as anomalous by exceeding the change significance level, in this case 5%.

The fine-grained CDF tail adjustor parameter $j$ was found to be a good detector of aberrant variability and used to assist in identifying broad mid-range platform issues. This parameter effectively functioned as ***"the canary in the mine"*** and was used to avoid significant corporate outage events.

Resolving system transaction performance profiles to a standard CDF with identical form and bounds between 0 and 1 enabled absolute comparison of $k$ values between system coarse-grained scalability. Additionally, it allowed performance profile comparison between systems with different transaction workload arrival rates. The relative value of the coarse grain parameter $k$ when comparing systems was an indicator of relative scalability with changing workloads. Table 3 shows these values for the 2 applications analyzed in this paper. This comparative concept needs more research to be confirmed more broadly.

A key aspect of this detection method was that the performance anomaly detection technique was platform and configuration agnostic. It did not rely on customization to a specific n-tier application architecture but was based on a whole of service monitoring approach from just in front of the servicing entity for each application being measured.

The shape of the change in $k$ and $j$ probability distributions gave a good indication of a system's variability in performance behavior. A tendency to grouping around the change of $-0.4 \leq \Delta \leq +0.4$ for the mainframe system indicated an application that was more tuned and subject to less variability in performance compared to the credit card system which had a more extended range $-0.6 \leq \Delta \leq 0.6$ . This needs more research to confirm more broadly.

## VI. Limitations of Detection Method

The following limitations of the detection method were identified:

TABLE 3: RELATIVE VALUES OF $k$ & $j$ FOR DIFFERENT APPLICATIONS

| *Application Service* | *Normal workload Values & Times* | *App service entities* | *Avg k during normal w'load* | *Avg j during normal w'load* |
|---|---|---|---|---|
| M/F Product system | 16-20,000 Tx/ 5min | Only one | 0.0083 | 0.2768 |
| Credit Card Payment system | 200-300 Tx/ 5min | Member 1 of 2 | 0.0037 | 0.0271 |

1. There needed to be enough sample points to properly fit the CDF curve and reduce variability.

2. Detecting changes in $k$ and $j$ meant that slow downs were detected relative to current application state. If the application was tuned well, then the method may detect events that are not practical to invest time to tune. Alternatively, if the system was in a poor state of repair there could be many slow down events that were missed while efforts were spent dealing with more significant slow down events. This is both a limitation and a benefit in that it naturally prioritizes the requirement to address problems.

3. Quantizing the probability segments into 0.1 lots and choosing the 5% significance level were arbitrary choices that may have de-sensitized the method. Research needs to be done on trending of these choices to fine tune the detection method.

4. A balancing consideration is the need for quick and light CPU resource consumption when performing automated regression calculations. The choice of regression engine performance and calculation structure (e.g. max number of regression iterations) needs to be considered.

## VII. Conclusions and Future Work

The hypothesis of this paper was that curve fitting using a CDF general form involving an exponential term that mimics the tail was an accurate method of representing a systems performance and its scalability. Further this modeling method could be used to detect application slow down events. The results showed that this hypothesis was true for the corporate context analyzed and a full day sample from 8am to 6pm Monday 4/2/2013. The technique was shown to work for a diverse range of workloads ranging from 76 Tx/5min to 19,025 Tx/ 5min. The method did not rely on customizations specific to the n-tier architecture of the systems being analyzed and so the performance anomaly detection technique was shown to be platform and configuration agnostic.

Further research is required to understand trending of relative values of $k$ and $j$ for applications with different transaction mixes and 3 tier application architecture footprints.

It is possible that the relationship between workload and $k$ can be used to track application scalability over time. The time of day view of changes in $k$ and $j$ showed a near linear relationship once the application had reached its stable transaction profile between 9am and 5pm depending on system. A further subject of study will be to investigate this relationship. A simplified approach might be to analyze the gradient of the line of best fit over an extended period as a useful indication of trends in scalability gain/loss. If this simple relationship correlates well then changes in gradient could be used to direct application tuning work and inform systems managers of the current risk position of applications.

The CDF form of the nonlinear function used for regression was univariate. This paper focused on the relationship between CDF and application service time for each sample over time to characterize the application performance signature. The service time was known to be impacted by the workload arrival rate and hence this variable included possible effects of changing workload arrival rates. A future piece of work involves examining multi-variate regression using arrival rate and service time. This technique may achieve a more sensitive alignment between the fitted curve and the regression curve. This would mean the change profile and hence anomaly detection capability could be improved.

The sample size used in this paper was a whole of day. A typical busiest day of the week was chosen for all commercial systems discussed in this paper – Mondays. This mechanism of application profiling will be productionized and longer term trending of the regression parameters examined. The hypothesis of further research work will be that trends in these parameters will reveal scalability trends that identify longer term performance issues due to system growth, need for tuning of dbase query plans, systemic growth in resource consumption and/or contention. In this way this detection technique can be analyzed for its ability to function in the short and long term time scales.

This detection technique could be combined with auto provisioning models such as those discussed in Tan, et al. [8], Stewart, et al. [9], and Tan, et al. [10] to achieve just in time performance anomaly prevention.

## ACKNOWLEDGMENT

Richard Gow would like to thank the following people for assisting him with retrieving transaction logs for the systems analyzed – Eduardo Soares, Neil Soutar, Peter Jadrijevic, and Matthew White.